# Wavevector Selective Metasurfaces and Tunnel Vision Filters

Vassili. A. Fedotov[1,*], Jan Wallauer[2], Markus Walther[2], Mauro Perino[1,3], Nikitas Papasimakis[1] and Nikolay I. Zheludev[1,4]

[1] *Optoelectronics Research Centre and Centre for Photonic Metamaterials, University of Southampton, SO17 1BJ, UK*
[2] *Department of Molecular and Optical Physics, University of Freiburg, D-79104 Freiburg, Germany*
[3] *Department of Information Engineering, University of Padova, Italy*
[4] *Centre for Disruptive Photonic Technologies, Nanyang Technological University, 637371 Singapore*

*\* e-mail: vaf@orc.soton.ac.uk*

Keywords: planar metamaterials; metasurfaces; metafilms; wavefront manipulation; flat optics

**Metasurfaces offer unprecedented flexibility in the design and control of light propagation, replacing bulk optical components and exhibiting exotic optical effects. One of the basic properties of the metasurfaces, which renders them as frequency selective surfaces, is the ability to transmit or reflect radiation within a narrow spectral band that can be engineered on demand. Here we introduce and demonstrate experimentally in the THz domain the concept of wavevector selective surfaces – metasurfaces transparent only within a narrow range of light propagation directions operating effectively as tunnel vision filters. Practical implementations of the new concept include applications in wavefront manipulation, observational instruments, vision and free-space communication in light-scattering environments, as well as passive camouflage.**

**INTRODUCTION**

Metasurfaces (also known as planar metamaterials or metafilms) are a special low-dimensional class of artificially structured media. It is represented by thin metal films and surfaces periodically patterned on a sub-wavelength scale, which can be readily fabricated using the existing planar technologies. Apart from their spectral selectivity [1] metasurfaces have demonstrated intriguing electromagnetic effects such as asymmetric transmission [2, 3] and optical activity without structural chirality [4]. They can exhibit resonant dispersion mimicking electromagnetically-induced transparency and slow light phenomenon [5, 6, 7], be invisible [8], efficiently convert polarization [9, 10, 11] or perfectly absorb radiation [12, 13]. Metasurfaces with gradient structuring anomalously reflect and refract light [10, 14, 15] and can act as lenses, wave-plates and diffraction gratings [16, 17, 18, 19]. Planar metamaterials are also able to enhance the light-matter interaction facilitating sensing [20], energy harvesting [21] and coherent radiation [22, 23].

The functionality of the most common types of planar metamaterials is determined by the individual resonant response of their basic structural elements – metamolecules, which are only weakly coupled to each other. When electromagnetic coupling between the metamolecules is strong [24] the relative phase of their excitation becomes important and the resulting spectral response is no longer determined by the individual resonances of the metamolecules. The metamaterial spectrum will be shaped by the collective, spatially coherent modes of metamolecular excitations that engage a large ensemble of metamolecules [25]. The introduction of structural disorder in such ensemble reduces the degree of coherency and lead to the weakening and broadening of its collective resonant response, which might be seen to vanish at even moderate levels of the disorder [24, 26]. Strong inter-metamolecular coupling is also responsible for the 'size effect' where the resonant transmission band of a metamaterial narrows with increasing size of the metamaterial sample [27].

In this paper we show that strong inter-metamolecular coupling can lead to a new phenomenon of 'tunnel vision', which renders a coherent metasurface as transparent within a very narrow range of light propagation directions. This effect is accompanied by 'rectification' of incident wavefronts, when initially spherical waves emerge as planar while traversing such metasurface in the absence of any spatial phase modulation or adaptive feedback [28]. The transmitted wavefronts appear parallel to the plane of the metamaterial and the effect does not depend on the curvature of the incident wavefronts.

As illustrated in Fig. 1, such response is fundamentally different from that of conventional convex lenses and recently demonstrated metasurface-based lenses [16, 19]. Indeed, a lens introduces a spatially dependent delay in the optical path, which compensates the curvature of the incident spherical wavefront if its source is located in the 'focal spot' of the lens. In a convex glass lens, for example, the spatial variation of optical delay is achieved by gradually reducing the thickness of the lens towards its edges (see Fig. 1A). In a metasurface-based lens the same is accomplished by changing the size and/or shape of its metamolecules: the resulting position-dependent phase lag for light scattered by the metamolecules mimicks spatial variation of optical delay in the glass lens (see Fig. 1B). For spherical waves originating at its focal point the lens converts the entire range of incident wavevectors into a much narrower range that converges around the optical axis of the lens.

The 'tunnel vision' effect can be understood as wavevector filtering, which occurs in a narrow transparency window of the metasurface corresponding to the collective resonance of its strongly coupled metamolecules. Indeed, in a regular planar array of identical metamolecules with the sub-wavelength period $d$ the coupling is strongest when the metamolecules are excited by a normally incident plane wave, i.e. they all oscillate in-phase. Any plane wave with its *k*-vector deviating from the array's normal (so that $k_\parallel \neq 0$) introduces a phase delay in the excitation of the metamolecules that linearly varies along the metasurface. The resulting de-phasing reduces the strength of coupling through the factor $\cos(k_\parallel d)$ and hence changes the energy of the collective mode [see Supplementary Information]. Consequently, the transparency window shifts to a different frequency and the metasurface becomes opaque. Operating at the frequency of its collective resonance the structure therefore acts as a wavevector selective surface (WSS): it admits only those partial plane waves whose *k*-vectors are parallel (or nearly parallel) to its normal, while all other waves are reflected back and/or absorbed (see Fig. 1C).

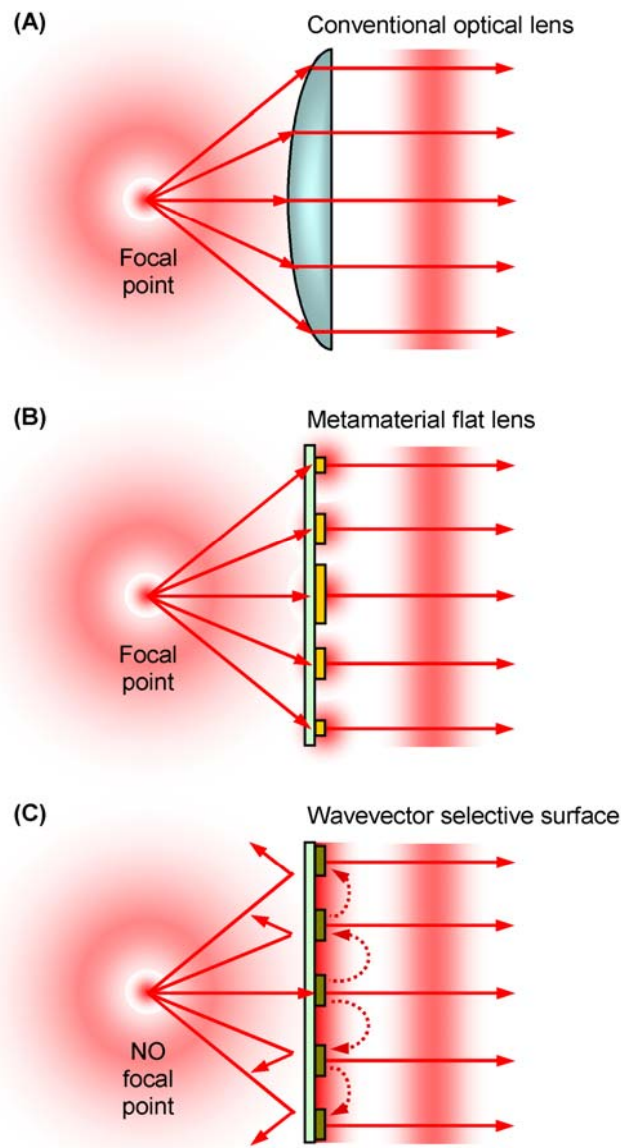

**Figure 1. Three ways of producing planar wavefronts with a point light source.**

**MATERIALS AND METHODS**

**Modeling.** The propagation of spherical wavefronts through a wavevector selective surface was simulated using 3D Maxwell's equations solver of the COMSOL Multiphysics simulation package 3.5a. The metasurface was modeled as a 14 × 14 array of asymmetrically split rings with the period $d = 565$ μm. The rings had the radius $r = 245$ μm and were split into wire sections $140^o$ and $160^o$ long separated by equal gaps. The wire sections had the width of 40 μm and were modeled as perfect electric conductors. The thickness of the supporting substrate was 120 μm and its permittivity was assumed to be $\varepsilon = 2.2$. The size of the simulation domain allocated for the transmitted wavefronts was 7910 μm × 7910 μm × 7910 μm.

**Experiment.** The metamaterial sample had been fabricated by etching a 9 μm thick copper layer covering one side of a 120 μm thick low-loss teflon substrate and closely resembled the modeled split-ring array both in terms of its size and design parameters. The spherical waves were produced by illuminating a 1750 μm large pinhole placed 500 μm away from the sample with a focused THz beam. The transmitted wavefronts were visualized using state-of-the-art THz-field imaging technique [29], which enabled mapping of electric-field component of the propagating waves with the spatial resolution of 160 μm and frequency resolution of 0.01 THz.

**RESULTS AND DISCUSSION**

As an example of WSS we consider here a planar metamaterial based on asymmetrically-split rings (ASR), a regular array of identical metamolecules formed by pairs of metallic arcs of different length (see inset to Fig. 2). For normally incident plane waves with E-field parallel to the splits its transmission spectrum features a narrow asymmetric pass-band with a sharp roll-off, which in our case is centered around $v_0 = 0.165$ THz (see Fig. 2A). This spectral feature corresponds to the resonant excitation of anti-symmetric charge-current mode, the so-called trapped or sub-radiant mode [30, 31], when charges $q$ and currents **j** induced in the opposite arcs of each ASR-metamolecule oscillate with equal amplitudes but opposite phases (see inset to Fig. 2). Such mode can be represented by a combination of an oscillating magnetic dipole with its moment being orthogonal to the metamaterial plane $\mathbf{m} = (0, 0, m_z)$, and an electric quadrupole **Q** characterized by two non-zero in-plane components $Q_{xy} = Q_{yx}$. Being arranged in a 2D-lattice with a sub-wavelength unit cell these multipoles cannot contribute to the far-field scattering of the array when they oscillate in-phase: coherent superposition of their fields results in electro-magnetic modes with the characteristic wavenumbers larger than $2\pi/\lambda_0$, and thus the multipole radiation by the array is trapped in the near-field zone. The absence of scattering in the far-field zone renders the metamaterial transparent, while the accumulation of energy in the spatially coherent surface waves ensure strong inter-metamolecular coupling [27].

For the ASR-metasurface the effect of wavevector filtering is particularly pronounced near 0.165 THz, at the sharp edge of its transparency band where a small blue shift of the band translates into a strong reduction of the metamaterial transmission. Such a shift corresponds to an increase of the trapped-mode's energy and therefore should result from the weakening of attractive inter-metamolecular coupling. In the array of ASR-metamolecules it is mediated by electric quadrupole-quadrupole interaction, and at oblique incidence it is affected by TM-component of the plane wave [see Supplementary Information].

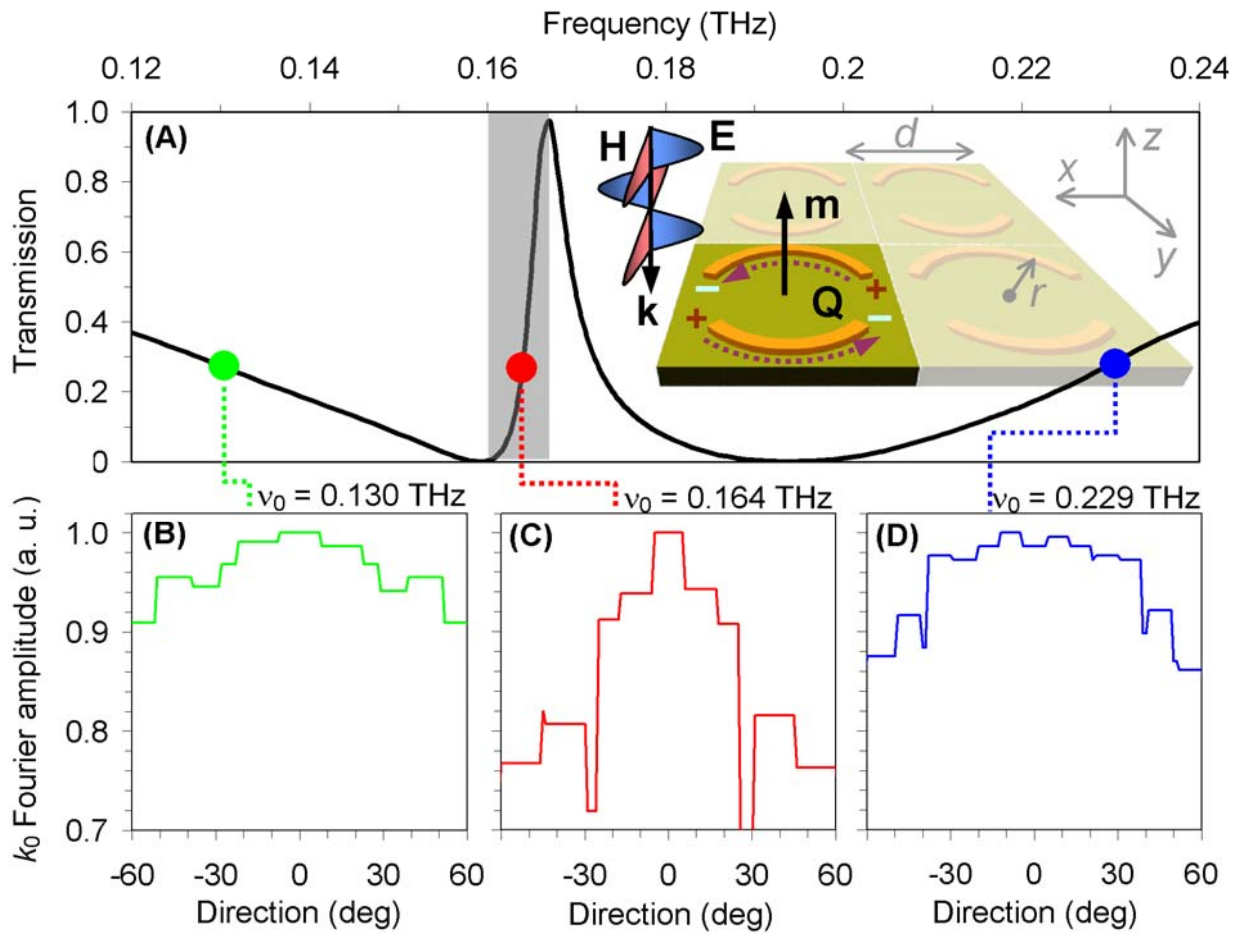

**Figure 2. Terahertz wavevector selective surface. (A) – WSS transmission spectrum for plane-wave illumination. Gray shading indicates the slope of the trapped-mode resonance. Inset shows a fragment of the WSS, a planar array of asymmetrically split rings. (B), (C) & (D) – Angular spectra of the transmitted wavevectors calculated for spherical-wave illumination at 0.130, 0.164 and 0.229 THz respectively.**

We first demonstrated the effect of wavevector filtering numerically, by simulating the propagation of spherical waves with a large wavefront curvature through the ASR-metasurface (see Figs. 3A-3C). The waves were produced by an electromagnetic point source placed close to the metasurface, at the distance equal to just one period of the ASR-array, $d$. The transmitted waves are visualized through the spatial variation of their phase and are presented for three characteristic frequencies $\nu_0$ (as indicated in Fig. 2A), which correspond to identical levels of transmission at the edges of the stop-band ($\nu_0 = 0.130$ THz and $0.229$ THz) and at the pass-band ($\nu_0 = 0.164$ THz). As evident from Figs. 3A and 3C, the curvature of the wavefronts at 0.130 and 0.229 THz remains practically unperturbed upon propagation through the metasurface. The situation changes markedly at the trapped-mode resonance, where the transmitted wavefront emerges nearly planar signifying thus the regime of wavevector filtering (see Fig. 3B).

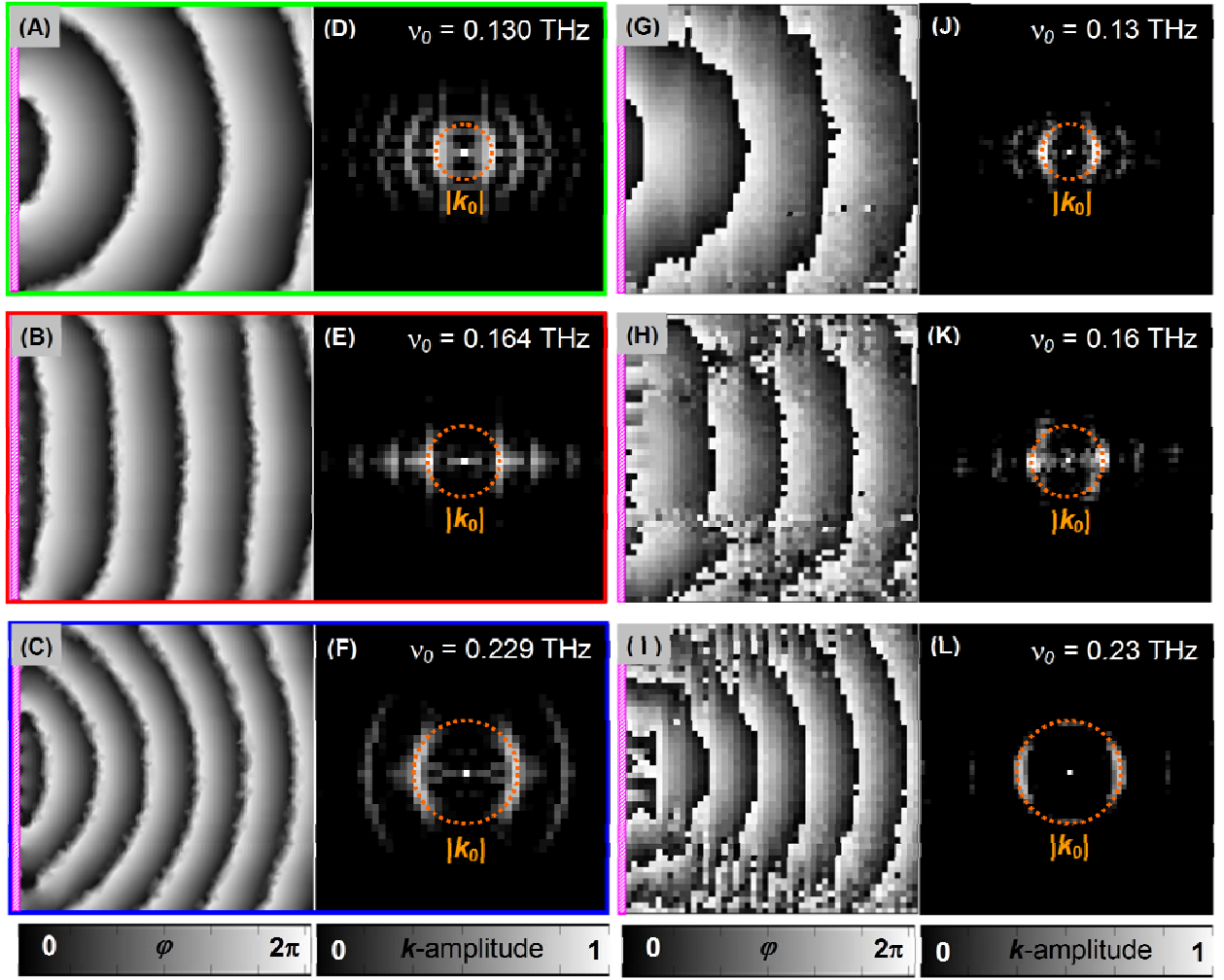

**Figure 3. Wavefront transformation by ASR-based wavevector selective surface.** (A - C) – Modeled spatial variations of the instantaneous phase, which visualize the wavefronts transmitted by the WSS at 0.130, 0.164 and 0.229 THz when it is illuminated by spherical waves. The waves originate on the left of the shaded purple bars, which mark the location of the WSS. (D - F) – Spatial-frequency spectra of the wavefronts shown in panels (A - C) respectively. (G - I) – Experimentally measured spatial variation of the instantaneous phase, which visualizes the wavefront of the initially spherical waves transmitted by WSS at 0.13, 0.16 and 0.23 THz. (J - L) – Spatial-frequency spectra of the wavefront patterns shown in panels (G - I), correspondingly.

To characterize the response of ASR-metasurface in terms of the transmitted $k$-vectors, i.e. partial plane-wave components, we plotted the spatial-frequency spectrum of the transmitted wavefront (see Figs. 3D-3F). A perfectly spherical wave would have been represented by a circle of the radius $k_0 = 2\pi \nu_0/c$ corresponding to the fundamental spatial frequency, as indicated by the doted circles in Figs. 3D-3F. Clearly, at $\nu_0 = 0.130$ THz and $\nu_0 = 0.229$ THz the spatial-frequency spectra have the form of partial concentric circles. The inner circle has the radius of $k_0$ and corresponds to the fundamental spatial frequency of the wavefront pattern. The outer circles represent higher harmonics $2k_0$, $3k_0$ etc. that are present due to unharmonic ('saw'-like) phase variations between 0 and $2\pi$. Given the extent of the circular patterns, the angles of incidence for the admitted partial plane-wave components fall in the relatively wide range from $-60°$ to $+60°$, which in our case is limited by the size of the modeled array. The amplitudes of the partial waves and therefore their relative contributions to the resulting wavefronts show little variation with the incidence angle, as evident from the angular spectra of the $k_0$-component presented in Figs. 2B and 2D. Thus, the response of the metasurface off the trapped-mode resonance is only weakly sensitive to the direction of the incident wavevectors.

At $\nu_0 = 0.164$ THz, however, the spatial-frequency spectrum collapses along the horizontal axis indicating that the transmission of the metasurface becomes $k$-dependent and most of the wavevectors deviating from the structure's normal are being rejected (see Fig. 3E). This is further illustrated in Fig. 2C, which shows that at $\nu_0 = 0.164$ THz the angular spectrum of the transmitted $k_0$-component converges along the $0^\circ$ direction.

The effect of wavevector filtering has been also confirmed experimentally using WSS-metamaterial sample that closely resembled the modeled ASR-array both in terms of its size and design parameters. The obtained images of the transmitted wavefronts are presented in Figs. 3G-3I. Evidently, the patterns of the wavefronts, as well as their spatial-frequency spectra plotted in Figs. 3J-3L show a very good agreement with our simulations. More experimental data, including the wavefront images obtained at other frequencies and their comparison with the results of our simulations can be found in Supplementary Information.

Unlike the wavevector manipulation performed by the lenses, the demonstrated principal of $k$-selectivity does not rely on gradient structuring. As the result, metamaterials with strong inter-metamolecular coupling can extract plane-wave components from arbitrary shaped wavefronts. A remote analogue of WSS functionality and the associated 'tunnel vision' effect might be found in conventional ray-optics systems such as astronomical telescopes: for the same magnification the telescopes with higher $f$-ratio (i.e. slower telescopes) will allow an observer to see stars and nebulas on a much darker background yielding overall higher contrast images. Such telescopes have smaller field of view, which limits the directions of the admitted light rays to those nearly parallel to the axis of the 'tunnel' (i.e. telescope) hence blocking most of the light scattered by the atmosphere and immediate surrounding.

Owing to their wavevector sensitive transmission response the WSSs could also be used as a simple passive means of concealing the presence of small objects under certain conditions. Observing a backlit object, one would normally collect both the fields radiated by the light source, as well as the fields scattered by the object. However, a WSS between the object and the observer will allow only the radiation from the source to reach the observer (provided the light source is far enough for the illumination to acquire a planar wavefront), whereas the non-normally incident scattered fields will be reflected by the WSS away from the observer. Such a scheme renders both the metamaterial screen and object practically undetectable, discerned from the background only through a small drop in the intensity due to the finite size of the screen (see illustration of the principle in Supplementary Information).

**CONCLUSIONS**

In conclusion, we have shown that strong electromagnetic coupling among the basic structural elements of a planar metamaterial provides a new degree of freedom to light manipulation, leading to an intriguing effect of wavefront rectification and tunnel vision. The effect results in arbitrary-shaped wavefronts becoming planar as they traverse the plane of the metamaterial in the absence of any spatial phase modulation or adaptive feedback, and is demonstrated here both theoretically and experimentally in the terahertz part of the spectrum. The proposed concept of wavevector selective surfaces can have a number of unique applications. For example, they can improve characteristics of observational instruments by blocking stray light and therefore acting as a flat analogue of a lens hood; or by reducing the effect of light scattering emanating from the immediate surrounding, dew, dust or scratches. WSSs can be exploited for directional filtering in free-space communications in highly turbid or strongly scattering media, yielding an improved signal-to-noise ratio. Finally, there is an exciting opportunity of using WSSs as a camouflage screen that provides a simple passive means to conceal the presence of small backlit objects.

ACKNOWLEGMENTS

This work is supported by the UK's Engineering and Physical Sciences Research Council through Career Acceleration Fellowship (V.A.F.) and Programme grant EP/G060363/1, by the Royal Society, and by the MOE Singapore grant MOE2011-T3-1-005.

# Supplementary Information for

# Wavevector Selective Metasurfaces and Tunnel Vision Filters

V. A. Fedotov[1,*], J. Wallauer[2], M. Walther[2], M. Perino[1,3], N. Papasimakis[1] and N. I. Zheludev[1,4]

[1] *Optoelectronics Research Centre and Centre for Photonic Metamaterials, University of Southampton, SO17 1BJ, UK*
[2] *Department of Molecular and Optical Physics, University of Freiburg, D-79104 Freiburg, Germany*
[3] *Department of Information Engineering, University of Padova, Italy*
[4] *Centre for Disruptive Photonic Technologies, Nanyang Technological University, 637371 Singapore*

[*] e-mail: **vaf@orc.soton.ac.uk**

## 1. Metamaterial Collective Resonance at Oblique Illumination

In this section we show, using a very simple model, how an obliquely incident plane wave may affect the energy of the collective sub-radiant mode of excitation of a planar metamaterial array.

We start with the well-known expression for the energy of the interaction between two parallel static magnetic dipoles **m₁** and **m₂** oriented such that the vector connecting the dipoles **r₁₂** = **r₂**-**r₁** is orthogonal to both **m₁** and **m₂**:

$$U_{12}^{\mathrm{m}} = \frac{\mu_0}{4\pi} \frac{m_1 m_2}{r_{12}^3}, \tag{S1}$$

here the positive sign of the expression indicates that magnetic dipole-dipole interaction is repulsive.

Such orientation of magnetic dipoles is encountered in the case of ASR-metamaterial at its trapped-mode resonance, where the dipole moments of its metamolecules at any given time are all oriented orthogonal to the plane of the metamaterial, and $r_{12} = d$ (see Fig. S1a). When induced by a normally incident plane wave the dipole moments acquire the same amplitude $m_z$ and phase factor $\cos(\omega t)$. Assuming that $d \ll \lambda$, the retardation effects can be neglected and the time-averaged energy of the interaction between two nearest metamolecules can be estimated using Eq. S1:

$$\langle U^{\mathrm{m}} \rangle \approx \frac{\mu_0}{4\pi} \frac{m_z^2}{d^3} \langle \cos^2(\omega t) \rangle = \frac{\mu_0}{4\pi} \frac{m_z^2}{2d^3}. \tag{S2}$$

At oblique incidence the magnetic dipoles may also become coupled to the plane wave through its magnetic field. For this to happen the wave must be TE-polarized so that it features a non-zero H-field component orthogonal to the metamaterial plane, $H_\perp$ (see Fig. S1a). The latter will naturally introduce a phase delay in the excitation of the dipoles $\mathbf{k}_\parallel \mathbf{r}$ (where $\mathbf{k}_\parallel$ is the component of the wavevector parallel to the metasurface), which will linearly vary along the metasurface and therefore result in a spatial modulation of the dipole moments with $\cos(\omega t - \mathbf{k}_\parallel \mathbf{r})$ profile. The energy of magnetic dipole-dipole interaction in this case is given by

$$\langle U^m \rangle \approx \frac{\mu_0}{4\pi} \frac{m_z^2}{d^3} \langle \cos(\omega t - \mathbf{k}_\| \mathbf{r}_1) \cos(\omega t - \mathbf{k}_\| \mathbf{r}_2) \rangle =$$
$$= \frac{\mu_0}{4\pi} \frac{m_z^2}{d^3} \left\langle \frac{\cos(2\omega t - \mathbf{k}_\|(\mathbf{r}_1 + \mathbf{r}_2)) + \cos(\mathbf{k}_\|(\mathbf{r}_2 - \mathbf{r}_1))}{2} \right\rangle \propto \frac{m_z^2}{d^3} \cos(k_\| d) \quad . \tag{S3}$$

Similar analysis may be performed taking into account other non-radiating multipolar excitations that contribute to inter-metamolecular interactions. In particular, one derives that the energy of electric quadrupole-quadrupole interaction, which also has a prominent presence in the ASR-metamaterial, will be proportional to $\cos(k_\| d)$ if the incident wave is TM-polarized:

$$\langle U^Q \rangle \propto -\frac{Q_{xy}^2}{d^5} \cos(k_\| d). \tag{S4}$$

Here $Q_{xy}$ (and $Q_{yx}$) are two non-zero components of the induced quadrupole moments, which are spatially modulated along the metasurface through an in-plane E-field component of the TM-wave, $E_\|$ (see Fig. S1b). The minus sign in the right side of Eq. S4 indicates that the quadrupole-quadrupole interaction in a uniformly excited metamaterial array is attractive.

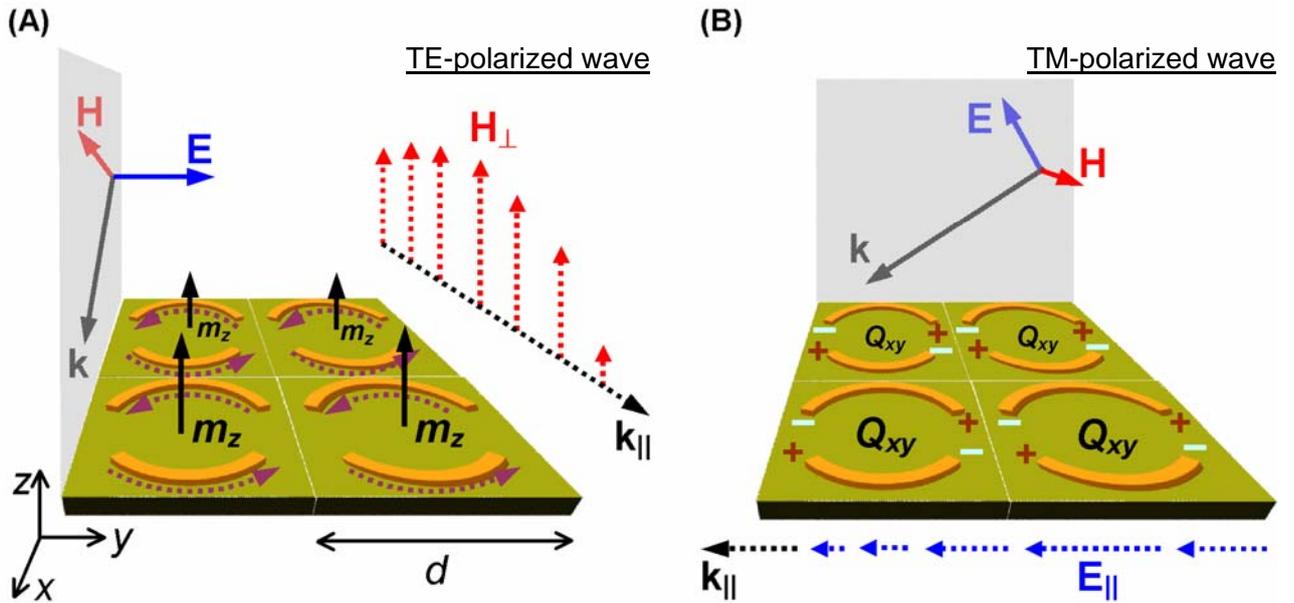

**Figure S1. Interaction of an obliquely incident plane wave with ASR-metamaterial. (A) – TE-polarized wave couples to the magnetic dipolar excitations in the ASR-metamolecules via its orthogonal magnetic field component $H_\perp$. (B) – TM-polarized wave couples to the electric quadrupolar excitations in the ASR-metamolecules via its longitudinal electric field component $E_\|$.**

Since the *cosine* function peaks at $k_\| = 0$, it follows from Eq. S3 that the energy of magnetic dipole-dipole interaction and correspondingly the energy of the collective trapped mode excited by TE-polarized wave would decrease as the angle of incidence increases ($k_\|$ increases). The opposite trend would be observed in the case of quadrupole-quadrupole interaction: due to initially attractive inter-metamolecular coupling the energy of the trapped mode excited by TM-polarized wave might only increase. The latter should result in a blue-shift of the transparency band associated with the trapped mode, which was confirmed by full-wave simulation of the metamaterial transmission using COMSOL 3.5a (see Fig. S2).

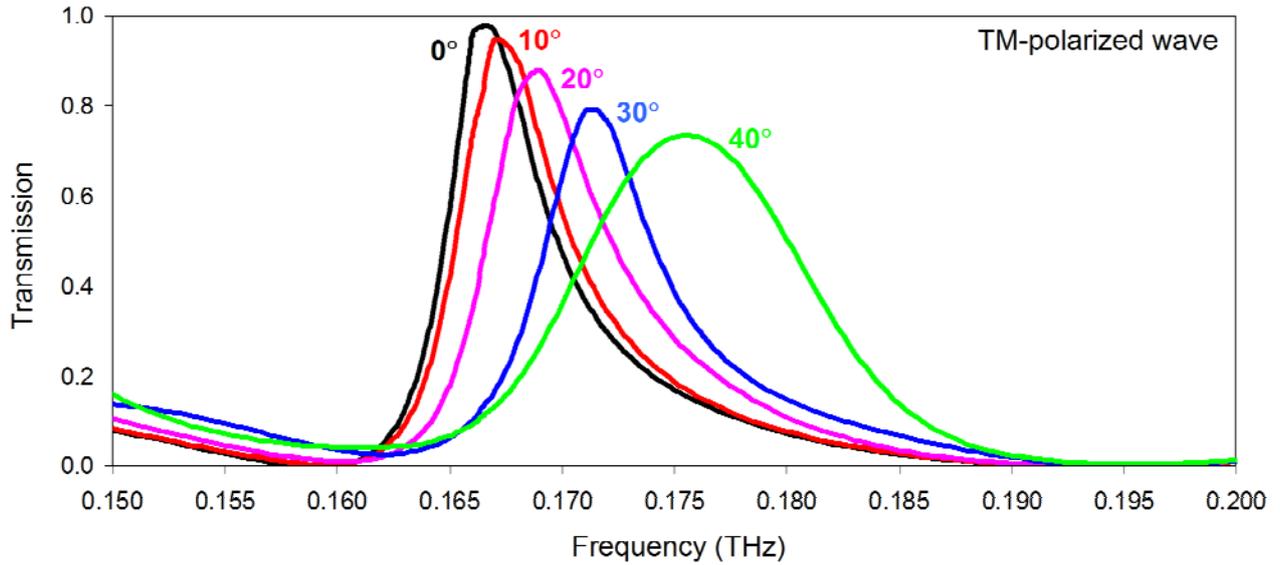

**Figure S2.** TM-wave transmission spectra of the ASR-metamaterial calculated for different angles of incidence.

## 2. Additional Experimental Data and Simulation Results

Below we provide the full set of experimentally obtained wavefront images, which are compared against the simulated ones. In the experiment, the spherical waves were produced by illuminating a 1750 μm large pinhole placed 500 μm away from the sample with a focused THz beam, as schematically shown in Fig. S3.

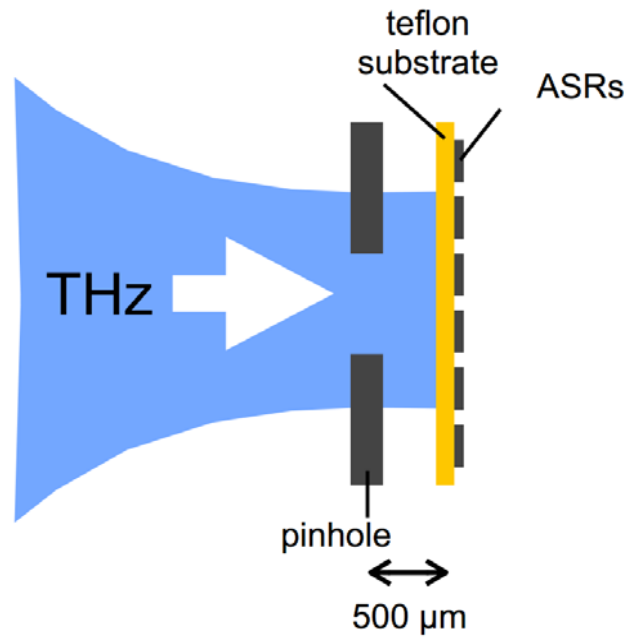

**Figure S3.** Schematic of the illumination arrangement used in the experiment.

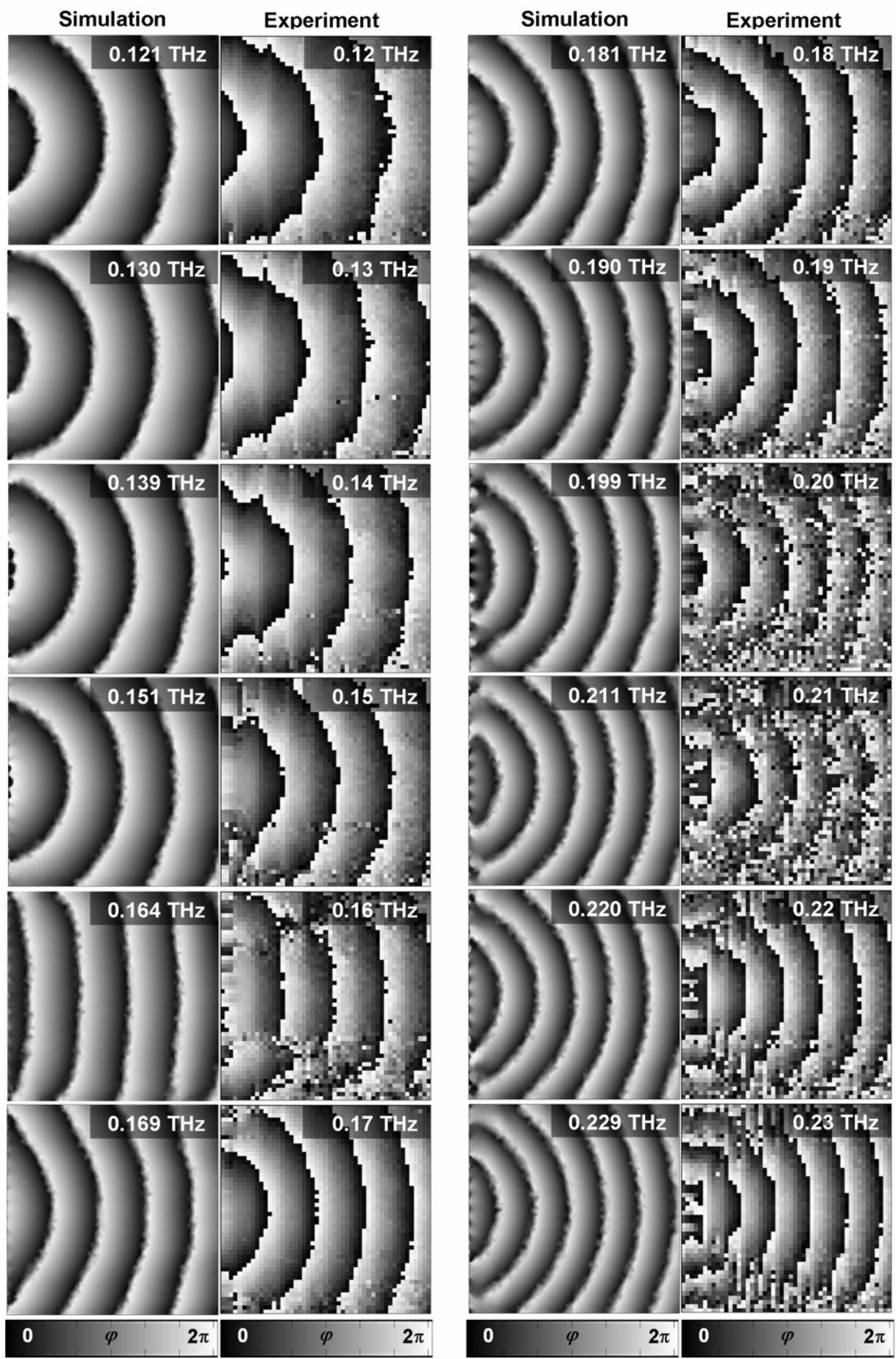

**Figure S4.** Initially spherical wavefronts visualized upon transmission through ASR-based WSS via the spatial variation of the phase, φ. The left side of the wavefront maps coincides with the plane of WSS.

Figure S5 shows far-field transmission spectrum of the ASR-metamaterial experimentally measured at normally incidence.

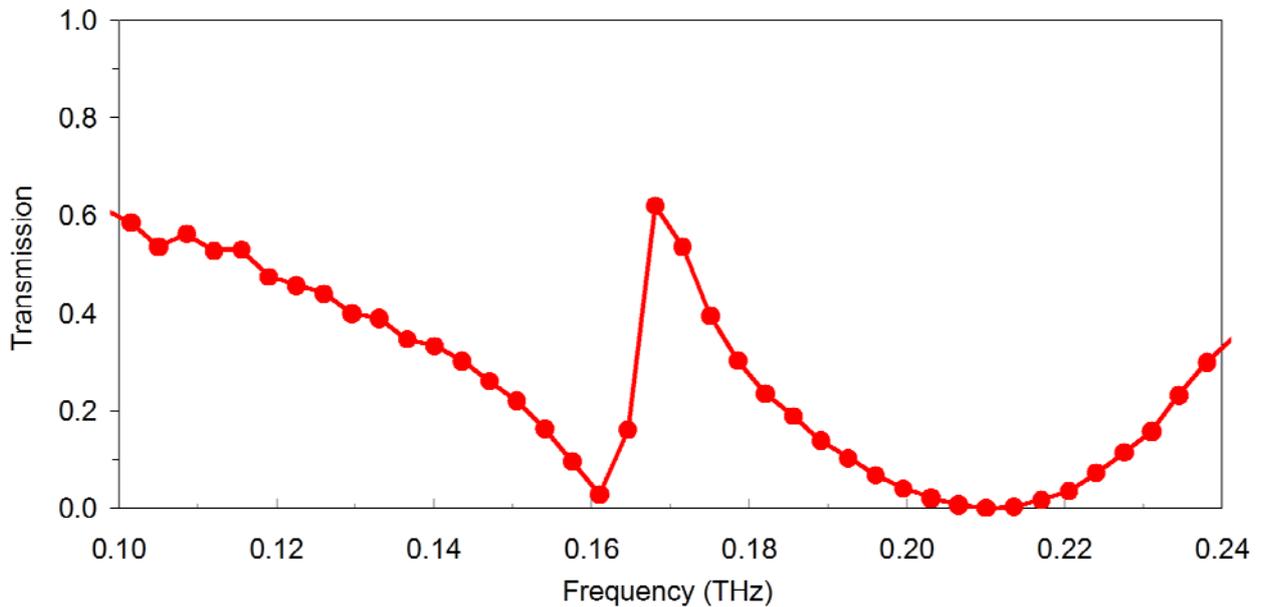

**Figure S5. WSS transmission spectrum experimentally measured using plane-wave illumination.**

**3. WSS as a Simple Alternative to Passive Camouflage**

In this section we illustrate the opportunity of using a WSS as a simple camouflage screen for a small backlit object and compare the principle with the conventional cloaking.

We consider here a specific case when the object to-be-concealed is illuminated from its back by the light sources, which are placed far enough for their radiation to reach the object as nearly plane waves. In conventional cloaking the waves will never scatter off the object and therefore reveal its presence as they are guided around the object by the cloaking shell (see Fig. S6A).

A WSS will conceal the object placed directly behind it in the following way. The light travelling from a distant scene reaches the WSS as plane waves incident nearly normal to its surface and therefore will be transmitted by the WSS forward virtually unperturbed. At the same time, the waves scattered by the object and propagating towards the observer will impinge onto the WSS along the directions deviating from its normal, and thus will be filtered out by the latter (see Fig. S6B). The reduction of overall transmission in this case can be made less obvious by using the metamaterial screen with the area substantially larger than the cross-section of the object.

A remote analogue of the described masking phenomenon might be encountered when one observes a very distant scene using a telescope and, at the same time, remains unaware of an object placed right in front of the telescope. As above, the darkening of the field of view in such case will be less obvious for a telescope with a larger aperture.

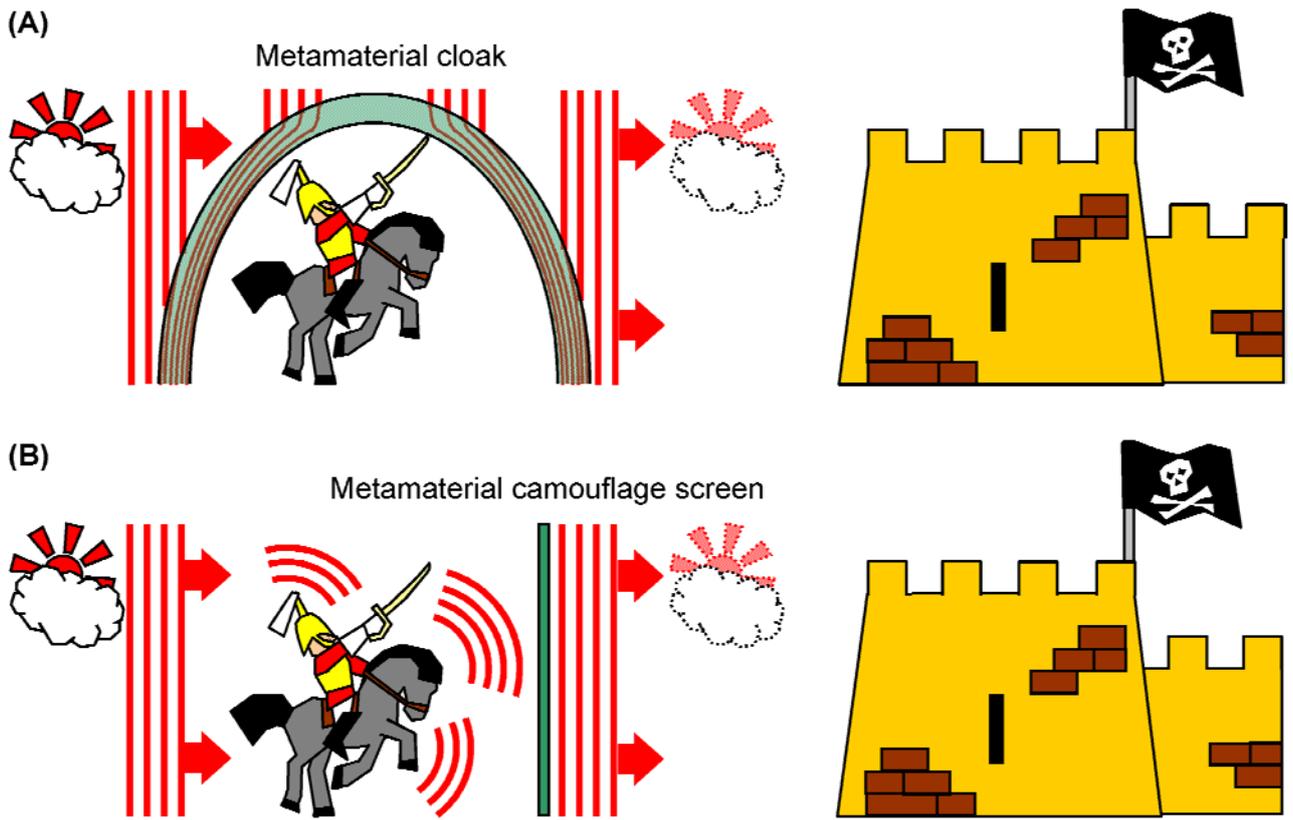

**Figure S6. Hiding backlit objects. (A)** – conventional 3D cloaking in action. The cloak is represented by green shell enclosing the dragoon. **(B)** – WWS acting as a camouflage screen (green bar in front of the dragoon).